\documentclass[aps,pre,superscriptaddress,showkeys]{revtex4-1}
\usepackage[T1]{fontenc}
\usepackage{graphicx}
\usepackage{geometry}

\begin{document}

\title{Mitigation of infectious disease at school:\\targeted class closure vs school closure}

\author{V. Gemmetto}
\affiliation{Data Science Laboratory, ISI Foundation, Torino, Italy}

\author{A. Barrat}
\affiliation{Aix Marseille Universit\'{e}, Universit\'e de Toulon, CNRS, CPT, UMR 7332, 13288 Marseille, France}
\affiliation{Data Science Laboratory, ISI Foundation, Torino, Italy}

\author{C. Cattuto}
\affiliation{Data Science Laboratory, ISI Foundation, Torino, Italy}

\begin{abstract}
\noindent \textbf{Background}\\
School environments are thought to play an important role in the community spread of airborne infections such as influenza because of the high mixing rates of school children. The closure of schools, therefore, has been proposed as an efficient mitigation strategy, with however high associated social and economic costs: hence alternative, less disruptive interventions are highly desirable. The recent availability of high-resolution contact networks in school environments provides an opportunity to design micro-interventions and compare the outcomes of alternative mitigation measures. \\

\noindent \textbf{Methods and Findings}\\
We consider mitigation measures that involve the targeted closure of school classes or grades based on readily available information such as the number of symptomatic infectious children in a class. We focus on the specific case of a primary school for which we have high-resolution data on the close-range interactions of children and teachers. We simulate the spread of an influenza-like illness in this population by using an SEIR model with asymptomatics, and compare the outcomes of different mitigation strategies. We find that targeted class closure affords strong mitigation effects: closing a class for a fixed period of time -- equal to the sum of the average infectious and latent durations -- whenever two infectious individuals are detected in that class decreases the attack rate by almost 70\% and significantly decreases the probability of a severe outbreak. The closure of all classes of the same grade mitigates the spread almost as much as closing the whole school.\\

\noindent \textbf{Conclusions}\\
Targeted class closure strategies based on readily available information on symptomatic subjects and on limited information on mixing patterns, such as the grade structure of the school, can be almost as effective as whole-school closure, at a much lower cost. This may inform public health policies for the management and mitigation of influenza-like outbreaks in the community.
\end{abstract}

\maketitle

\section{Introduction}
It has been long known~\cite{Longini:1982,Viboud:2004}
that children play an important role in the community spread of infectious disease, in particular of influenza.
The many contacts children have with one another at school increase their risk of being infected
by several airborne transmissible pathogens,
and make schools an important source of transmission to households,
from where the disease can spread further.
For instance, during the 2009 H1N1 pandemic, a correlation was observed~\cite{Chao:2010}
between the opening dates of schools and the onset of widespread transmission of H1N1 in the US.
Similarly, the timing of school terms, with the corresponding changes in contact patterns,
has been shown to explain the evolution of the H1N1 epidemic in the UK~\cite{Eames:2012}.

School closure is thus regarded as a viable mitigation strategy for epidemics~\cite{WHO,Cauchemez:2009},
especially in the case of novel pandemics for which pharmaceutical interventions, such as vaccines,
are not readily available and delaying disease spread is a priority.
The impact of school closure on the spread of infectious disease has been studied
using historical data~\cite{Heymann:2004,Cauchemez:2008,Cowling:2008,Rodriguez:2009,Jackson:2011,Earn},
comparison of contact patterns during week days, weekends and holiday periods~\cite{Hens:2009,Eames:2011,Eames:2012},
and agent-based models at different scales~\cite{Ferguson,Glass,Lee,Cauchemez:2008}.
School closure, however, comes with a steep socio-economic cost,
as parents need to take care of their children and might be forced to take time off work.
This can even have a detrimental impact on the availability of public health staff.
Such harmful side effects have led to question the effective benefit of school closure~\cite{Brown,Dalton:2008}
and prompt research on the design and evaluation of non-pharmaceutical low-cost mitigation strategies.

In this context, the availability of data on contacts between school children is a crucial asset on two accounts:
First, even limited information on mixing patterns within and between classes or grades
can suggest more refined strategies than whole-school closure.
Second, high-resolution contact data allow to develop individual-based computational models of disease spread
that can be used to test and compare different mitigation strategies.
Because of this, over the last few years a great deal of effort has been devoted
to gathering data on human contact patterns in various environments~\cite{Read:2012},
using methods that include diaries and surveys~\cite{Edmunds,Mossong,Read,Zagheni,Mikolajczyk,Hens:2009,Conlan,Smieszek:2012,Potter,Danon:2013},
and more recently wearable sensors that detect close-range proximity~\cite{Eagle:2006,Salathe,Hornbeck:2012}
and face-to-face contacts~\cite{SocioPatterns,Cattuto,Isella,Stehle:2011a,Isella2}.

In this study we use a high-resolution contact network
measured by using wearable sensors in a primary school~\cite{Stehle:2011a}.
The data show that children spend more time in contact with children of the same class
(on average three times more than with children of other classes) and of their own grade~\cite{Stehle:2011a}.
This is expected to be a rather general feature of schools,
due both to age homophily~\cite{McPherson} and schedule constraints,
and suggests that transmission events might take place preferentially within the same class or grade.
We thus consider targeted and reactive mitigation strategies in which
one class or one grade is temporarily closed whenever symptomatic individuals are detected.
To evaluate the effectiveness of such micro-interventions
we use our high-resolution contact network data~\cite{SocioPatterns,Stehle:2011a} 
to build an individual-based model of epidemic spread,
and we compare, in simulation, the performance and impact on schooling
of different targeted mitigation strategies with the closure of the whole school.

\section{Methods}

\subsection*{High-resolution contact network data}
We use a high-resolution contact network measured by the SocioPatterns collaboration~\cite{SocioPatterns}
using wearable proximity sensors in a primary school. The sensors detect the face-to-face
proximity relations (``contacts'') of individuals with a 20-seconds temporal resolution~\cite{Cattuto}.
The time-resolved contact network considered here, analyzed in Ref.~\cite{Stehle:2011a},
describes the contacts among $232$ children and $10$ teachers in a primary school in Lyon, France,
and covers two days of school activity (Thursday, October $1^{st} $ and Friday, October $2^{nd} $ 2009).
The school is composed by $5$ grades, each of them comprising two classes, for a total of $10$ classes.
Contacts events are individually resolved, and their starting and ending times
are known up to the 20-second resolution of the measurement system.

\subsection*{Extending the temporal span of the empirical data}

Realistic parameters for the infectious and latent periods of influenza-like disease are of the order of days.
Since the dataset we use only spans two school days, our numerical simulations will unfold
over time scales longer than the duration covered by the contact dataset. To address this problem,
several possibilities to extend in time the empirical contact data have been explored~\cite{Stehle:2011b}.
Here we consider a simple periodic repetition of the 2-day empirical data,
modified to take into account specific features of the school environment under study.
First, since our data only describes contacts during school hours,
we assume that children are in contact with the general community for the rest of the day.
Moreover, children in France do not go to school on Wednesday, Saturday and Sunday:
on these days, therefore, children are also considered in contact with the general community.
Overall the temporal contact patterns we use have the following weekly scheme:
\begin{itemize}
\item[i)] Monday and Tuesday correspond to the first and second day of the empirical dataset: between  8.30am and 5:00pm contacts within the school are described by the empirical data. Outside of this interval, children are assumed to be isolated from one another and in contact with the community.
\item[ii)] Wednesday: children are in contact with the community for the entire day.
\item[iii)] Thursday and Friday: the first and second day of the empirical dataset are repeated as in i).
\item[iv)] Saturday and Sunday: children are in contact with the community for the entire weekend.
\end{itemize}
The above weekly sequence is repeated as many times as needed.
Other extension procedures include partial reshuffling of the participants' identities
across days~\cite{Stehle:2011b}, to model the partial variability of each individual's contacts
from one day to the next. Here we limit our investigation to the simple scheme outlined above,
because a repetition procedure is appropriate to model a school environment,
where activities follow a rather repetitive daily and weekly rhythm,
and each child is expected to interact every day with approximately the same set of individuals,
namely the members of her/his class and her/his acquaintances in other classes.

\subsection*{Epidemic model} 
To simulate the spread of an influenza-like disease
we consider a stochastic SEIR model with asymptomatic individuals, 
with no births, nor deaths, nor introduction of individuals~\cite{Anderson}.
In such a compartmental model each individual at a given time can be in one of five possible states:
susceptible (S), exposed (E), infectious and symptomatic (I), infectious and asymptomatic (A), and recovered (R).
Whenever a susceptible individual is in contact with an infectious one,
s/he can become exposed at rate $\beta$ if the infectious individual is symptomatic,
and $\beta/2$ if the infectious individual is asymptomatic.
Exposed individuals, who cannot transmit the disease, become infectious with a fixed rate $\mu$,
where $1/\mu $ represents the average duration of the latent period.
Exposed individuals becoming infectious have a probability $p_A$
of being asymptomatic (A) and a probability $1-p_A$ of being symptomatic (I).
Both symptomatic and asymptomatic infectious individuals recover at a fixed rate $\gamma$
($1/\gamma$ is the average duration of the infectious period)
and acquire permanent immunity to the disease.

As mentioned above, our data describe human contacts only within the school premises. 
During the spread of an epidemic in the community, however, exposure to infectious individuals
also occurs outside of school. Accordingly, we consider that individuals
have a generic risk of being contaminated by infectious individuals outside of the school.
For simplicity, here we assume that this risk is uniform
and we introduce it into the model through a fixed rate of infection $\beta_{com}$.
That is, the probability that a susceptible individual, during a time interval $dt$,
becomes exposed due to random encounters outside of school is $\beta_{com} \, dt$. 

Finally, we assume that symptomatic individuals are detected at the end of each day.
They are subsequently isolated until they recover and therefore cannot transmit the disease anymore.
Asymptomatic individuals, on the other hand, cannot be detected and thus are not isolated.
Each simulation starts with a completely susceptible population,
except for a single, randomly chosen infectious individual,
chosen as symptomatic with probability $1-p_A$ and asymptomatic with probability $p_A$.

We consider the following parameter values for the SEIR model:
$\beta = 3.5 \cdot 10^{-4} \, s^{-1} $,
$\beta_{com} = 2.8 \cdot 10^{-9} \, s^{-1} $,
$1 / \mu = 2 $ days, 
$1 / \gamma = 4 $ days.
The fraction of infected asymptomatic individuals is set to $p_{A}= 1/3$. 
These parameter values are in line with those used to describe influenza-like illnesses~\cite{Tizzoni:2012}.

Moreover, we carry out the following sensitivity analyses:
First, we consider a larger value of $\beta_{com}$
while keeping fixed the values of the other parameters,
to investigate the role of the generic risk of infection in the community.
Second, we report in the Supplementary Text the results obtained
with two different sets of parameters
corresponding to faster spreading processes, namely:
(i) $\beta = 6.9 \cdot 10^{-4} \, s^{-1}$;
$\beta_{com} = 2.8 \cdot 10^{-9} \, s^{-1}$;
$1/ \mu = 1 $ day; 
$1/ \gamma = 2 $ days, 
$p_A=1/3$, and
(ii) $\beta = 1.4 \cdot 10^{-3} \, s^{-1}  $,
$\beta_{com} = 2.8 \cdot 10^{-9} \, s^{-1} $,
$1/ \mu = 0.5 $ day,
$1/ \gamma = 1 $ day, $p_A=1/3$.

\subsection*{Mitigation measures} 
The baseline mitigation measure is given by the isolation of symptomatic children at the end of each day.
We consider the three following additional strategies:
whenever the number of symptomatic infectious individuals detected in any class reaches a fixed threshold,
\begin{itemize}
\item[(i)] the class is closed for a fixed duration (``targeted class closure'' strategy);
\item[(ii)] the class and the other class of the same grade are both closed for a fixed duration  (``targeted grade closure'' strategy);
\item[(iii)] the entire school is closed for a fixed duration (``whole school closure'' strategy).
\end{itemize}
In all cases, the children affected by the closure are considered to be
in contact with the community during the closure period -- with the exception of detected infectious cases --
and therefore they have a probability per unit time $\beta_{com}$ of acquiring the disease.
When the closure is over, the class (or grade) is re-opened and the corresponding children go back to school.

For benchmarking purposes, in the Supplementary Text we also consider
strategies based on random class closures:
whenever the number of symptomatic infectious individuals
detected in any class reaches a fixed threshold, 
\begin{itemize}
\item[(iv)] one random class, different from the one in which symptomatic individuals are detected, is closed (``random class closure'' strategy)
\item[(v)] the class and a randomly chosen one in a different grade are closed (``mixed class closure'' strategy).
\end{itemize}

Note that during the course of an epidemic, in principle,
several classes can be closed at the same time or successively,
but once a class (or grade) is re-opened, we do not allow it to be closed again.
Similarly, when using the whole-school closure strategy,
we assume for simplicity that once the school is re-opened it cannot be closed again.

All of the closure strategies describe above depend on two parameters:
the closure-triggering threshold, i.e., the number of symptomatic individuals required to trigger the intervention,
and the duration of the closure.
We will explore thresholds of $2$ or $3$ symptomatic individuals
and closure durations ranging from $24$ to $144$ hours (from $1$ to $6$ days).
Closure durations are specified in terms of absolute time:
for instance, a $72$ hours closure starting on a Thursday night spans
the following Friday, Saturday and Sunday and ends on the next Monday morning.

\subsection*{Simulation and analysis} 
For each set of model parameters for the SEIR model, and for each set of parameters of every mitigation strategy
we simulate $5000$ realizations of the epidemic process.
We compare the performance of different strategies by measuring
(i) the fraction of stochastic realizations that yield an attack rate (fraction of individuals
affected by the disease) higher than $10\%$, and
(ii) the average number of final cases in the population.
We also quantify the burden of each strategy by computing the number of lost schools days.
defined by adding up the number of school days missed by each class affected by the intervention.
A closure of one class during a normal school day counts as $1$ lost day,
whereas the closure of the entire school counts as $10$ lost days, as there are $10$ classes in the school.
We do not count Wednesdays and week-ends spanned by the closure interval.

\section{Results}
Here we provide results corresponding to the parameters values
$\beta = 3.5 \cdot 10^{-4} \, s^{-1}$,
$\beta_{com} = 2.8 \cdot 10^{-9} \, s^{-1} $,
$1/ \mu = 2$ days,
$1/ \gamma = 4 $ days.  
The results for the two other sets of parameters
are qualitatively similar and are discussed in the Supplementary Text.

In Table~\ref{table:1} we report the fraction of stochastic realizations
that lead to an attack rate (AR) higher than $10 \%$,
for each mitigation strategy and each set of parameter values
of the strategy (closure triggering threshold and closure duration).
As a baseline, we also report the attack rate obtained when no closure is implemented,
i.e., when the only mitigation measure is the isolation of symptomatic individuals
at the end of each school day.
Even when no closure strategy is implemented
the majority of realisations ($65.4\%$) do not lead to a large outbreak.
The probability of a large outbreak is reduced by all of the closure strategies.
We observe a larger reduction for smaller values of the closure-triggering threshold
and for longer closure durations.
On closing whole grades ($2$ classes) rather than individual classes
we report a smaller percentage of realizations leading to large outbreaks.
\begin{table}
{\begin{center} 
\begin{tabular}{|c|c|c|c|}
\hline
Closure strategy & Targeted class & Targeted grade & Whole school  \\
(threshold, duration)  & & & \\
\hline \hline
No closure & 34.6 & 34.6 & 34.6  \\
\hline \hline
3, 24 h & 30.5 & 29.7 & 26.0  \\
\hline
3, 48 h & 28.1 & 23.5 & 23.2 \\
\hline
3, 72 h & 23.4 & 18.4 & 14.8 \\
\hline
3, 96 h & 23.5 & 20.3 & 13.0  \\
\hline
3, 120 h & 20.1 & 17.3 & 7.5 \\
\hline
3, 144 h & 19.7 & 16.3 & 5.6  \\
\hline \hline
2, 24 h & 28.6 & 27.0 & 22.9 \\
\hline
2, 48 h & 22.0 & 21.6 & 17.8 \\
\hline
2, 72 h & 17.4 & 16.2 & 14.4 \\
\hline
2, 96 h & 13.6 & 11.2 & 11.0  \\
\hline
2, 120 h & 10.2 & 7.2 & 3.2 \\
\hline
2, 144 h & 11.6 & 6.8 & 1.6 \\
\hline
\end{tabular}
\caption{Percentage of realizations leading to an attack rate higher than $10 \%$,
for different mitigation strategies and for various closure-triggering thresholds and closure durations.
Parameter values:
$\beta = 3.5 \cdot 10^{-4} \, s^{-1} $,
$ \beta_{com} = 2.8 \cdot 10^{-9} \, s^{-1} $,
$1/ \mu = 2 $ days, 
$1/ \gamma = 4 $ days,
$p_A = 1/3$.
}
\label{table:1}
\end{center}}
\end{table}

In Table~\ref{table:2} we complement the above results by reporting,
for each strategy and parameter choice,
the final number of cases (averages and confidence intervals)
for realizations leading to an attack rate larger than $10\%$.
For small enough closure triggering thresholds and long enough closure durations,
all closure strategies achieve a strong reduction of the final epidemic size.
Strategies affecting more classes also have a stronger effect,
but in those cases we observe large confidence intervals
and large overlap of the epidemic sizes for different choices
of the strategy parameters.
In particular, for small closure-triggering thresholds
the targeted class and targeted grade strategies yield reductions
in the number of large outbreaks that are similar to those observed
for the closure of the whole school.
\begin{table}
{\begin{center} 
\begin{tabular}{|c|c|c|c|}
\hline
Closure strategy & Targeted class & Targeted grade & Whole school  \\
(threshold, duration)  & & & \\
\hline \hline
No closure & 179 [149,203] & 179 [149,203] & 179 [149,203]  \\
\hline \hline
3, 24 h & 162 [122,199] & 166 [112,196] & 170 [151,202]  \\
\hline
3, 48 h & 135 [48,197] & 138 [40,188] & 162 [43,199] \\
\hline
3, 72 h & 101 [33,186] & 103 [30,177] & 146 [28,198] \\
\hline
3, 96 h & 92 [29,184] & 88 [26,169] & 120 [27,195]  \\
\hline
3, 120 h & 75 [29,170] & 62 [25,163] & 67 [26,192] \\
\hline
3, 144 h & 71 [26,168] & 58 [24,161] & 55 [25,180]  \\
\hline \hline
2, 24 h & 165 [91,195] & 170 [141,199] & 173 [139,198] \\
\hline
2, 48 h & 124 [32,179] & 142 [35,191] & 170 [62,199] \\
\hline
2, 72 h & 96 [30,170] & 113 [29,180] & 149 [48,201] \\
\hline
2, 96 h & 75 [27,152] & 94 [26,184] & 141 [31,196]  \\
\hline
2, 120 h & 69 [25,140] & 73 [25,181] & 133 [30,195] \\
\hline
2, 144 h & 51 [27,111] & 52 [26,138] & 57 [25,192]  \\
\hline
\end{tabular}
\caption{Average final number of cases, computed for realizations leading to an attack rate higher than $10\%$.
In square brackets we provide the $5^{th}$ and $95^{th}$ percentiles.
Parameter values:
$\beta = 3.5 \cdot 10^{-4} \, s^{-1}$,
$\beta_{com} = 2.8 \cdot 10^{-9} \, s^{-1}$,
$1/ \mu = 2 $ days, 
$1/ \gamma = 4 $ days,
$p_A = 1/3$.
}
\label{table:2}
\end{center}}
\end{table}

\begin{figure}[htbp]
\begin{center}
\includegraphics[scale=0.45]{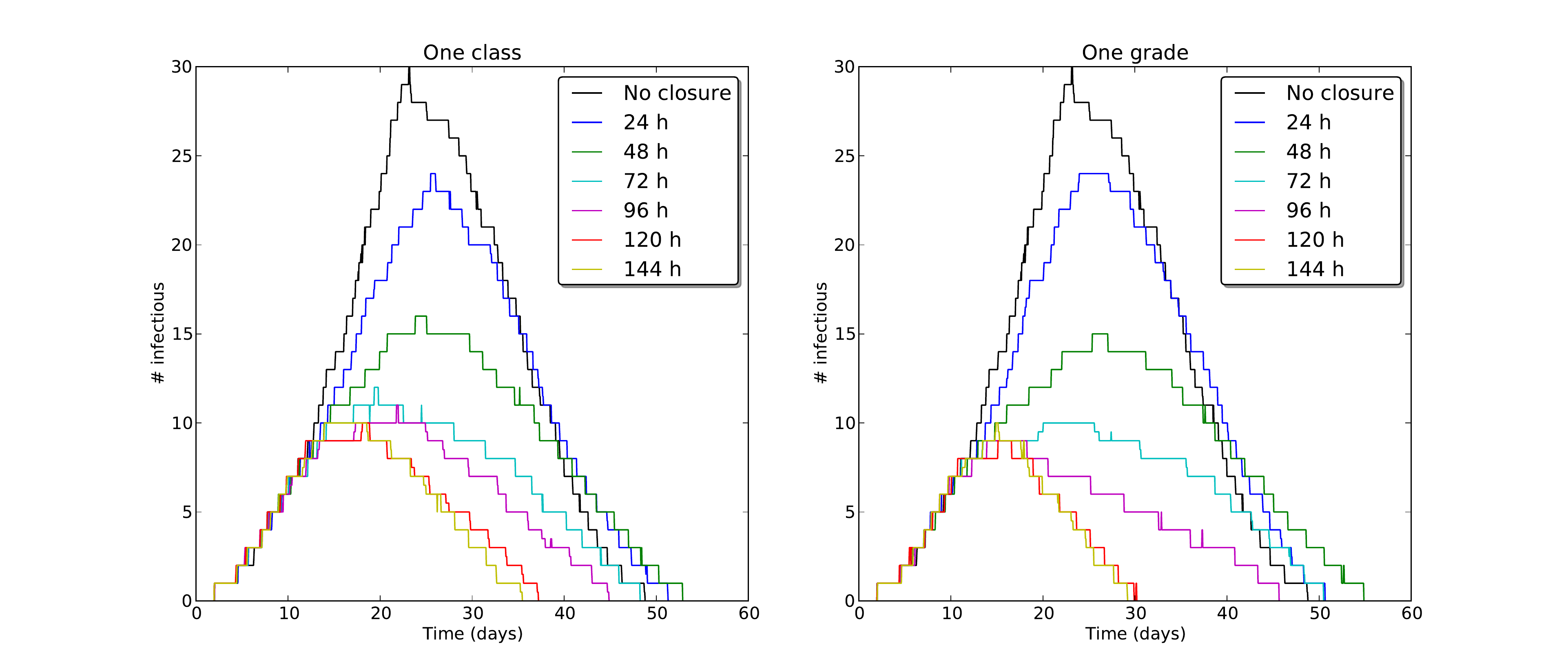}
\end{center}
\caption{Temporal evolution of the median number of infectious individuals for several closure durations,
at a fixed closure-triggering threshold of $3$ symptomatic cases.
Left: targeted class closure. Right: targeted grade closure.
Only runs with an attack rate (AR) higher than $10\%$ are taken into account.
Parameter values:
$\beta = 3.5 \cdot 10^{-4} \, s^{-1}$,
$\beta_{com} = 2.8 \cdot 10^{-9} \, s^{-1}$,
$1 / \mu = 2 $ days, 
$1/ \gamma = 4 $ days,
$p_A = 1/3$.
\label{fig:thr3_add2}
}
\end{figure}

\begin{figure}[htbp]
\begin{center}
\includegraphics[scale=0.55]{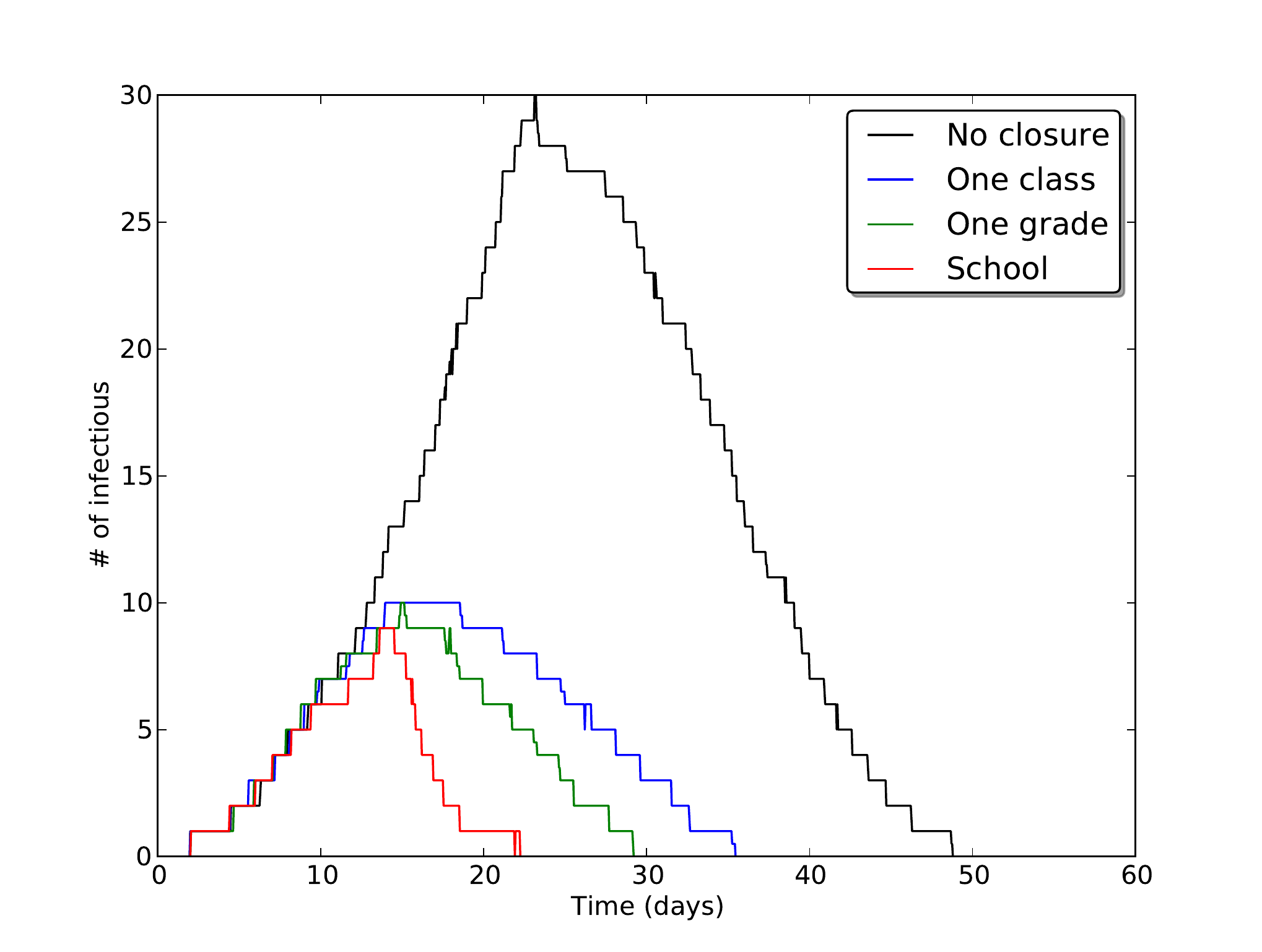}
\end{center}
\caption{Temporal evolution of the median number of infectious individuals
for the targeted class, targeted grade, and whole school closure strategies,
at a fixed closure-triggering threshold of $3$ infectious individuals
and closure duration of $144$ hours ($6$ days).
The no-closure scenario is provided for reference.
Only realizations with an attack rate (AR) higher than $10\%$ are taken into account.
Parameter values:
$\beta = 3.5 \cdot 10^{-4} \, s^{-1}$,
$\beta_{com} = 1.4 \cdot 10^{-8} \, s^{-1}$,
$1/ \mu = 2 $ days,
$1/ \gamma = 4 $ days,
$p_A = 1/3$.
\label{fig:th3_72h_add2}
}
\end{figure}

Figures \ref{fig:thr3_add2} and \ref{fig:th3_72h_add2}
display the the temporal evolution of the median number of infectious individuals
for several mitigation strategies, when only realizations leading to an attack rate higher than $10\%$ are considered.
Figure~\ref{fig:thr3_add2} shows the effect of closure duration for the targeted class and targeted grade strategies
at a fixed closure-triggering threshold of $3$ symptomatic cases.
Longer closures lead to shorter and smaller epidemic peaks.
Closure durations of $5$ or $6$ days ($120$ and $144$ hours, respectively) lead to very similar epidemic curves.
Figure~\ref{fig:th3_72h_add2}, on the other hand,
compares the epidemic curves for the targeted class, targeted grade and whole school strategies
at fixed closure-triggering threshold and closure duration parameters.
The targeted class closure strategy already yields a large reduction of the epidemic peak,
and this reduction is only slightly improved by the targeted grade closure
and whole school closure strategies (for the same closure durations).

Finally, in Table~\ref{table:4} we report  the impact on the schooling system of each closure strategy,
quantified by the average number of lost school days aggregated over all affected classes.
In all cases, a high percentage of realizations lead to zero impact,
corresponding to (i) situations in which the outbreak stays confined
and the closure-triggering threshold is never reached in any class,
or (ii) to cases in which the closure happens on days
during which the school is scheduled to be closed (Wednesdays or week-ends).
Whenever schooldays are effectively lost,
we observe a much greater impact for the whole school closure
than for the alternative strategies of closing one class or one grade only.
\begin{table}
{\begin{center} 
\begin{tabular}{|c|c|c|c|}
\hline
Closure strategy & Targeted class & Targeted grade & Whole school  \\
(Threshold, duration)  & & & \\
\hline \hline
3, 24 h & 2.14 (3.27) [0-9] & 2.34 (3.58) [0-10] & 2.50 (4.33) [0-10]  \\
\hline
3, 48 h & 3.04 (4.83) [0-13] & 3.00 (5.12) [0-14] & 4.42 (7.26) [0-20] \\
\hline
3, 72 h & 3.01 (5.17) [0-15] & 3.21 (5.44) [0-16] & 5.38 (8.13) [0-20] \\
\hline
3, 96 h & 4.49 (7.62) [0-22] & 4.93 (8.21) [0-24] & 8.10 (11.5) [0-30]  \\
\hline
3, 120 h & 4.50 (8.41) [0-26] & 5.20 (9.06) [0-28] & 9.38 (13.4) [0-40] \\
\hline
3, 144 h & 4.67 (8.72) [0-27] & 5.33 (9.41) [0-28] & 9.60 (13.7) [0-40] \\
\hline \hline
2, 24 h & 2.18 (3.36) [0-9] & 2.31 (3.46) [0-10] & 3.38 (4.73) [0-10] \\
\hline
2, 48 h & 2.42 (4.37) [0-13] & 3.05 (4.77) [0-14] & 4.60 (7.16) [0-20] \\
\hline
2, 72 h & 2.57 (4.77) [0-15] & 3.44 (5.45) [0-16] & 7.12 (8.54) [0-20] \\
\hline
2, 96 h & 3.14 (6.11) [0-20] & 3.92 (6.80) [0-22] & 8.64 (11.2) [0-30]  \\
\hline
2, 120 h & 3.55 (6.37) [0-18] & 4.38 (7.68) [0-22] & 9.32 (12.4) [0-30] \\
\hline
2, 144 h & 4.32 (7.38) [0-22] & 4.63 (7.61) [0-22] & 11.54 (13.7) [0-40] \\
\hline
\end{tabular}
\caption{
Number of lost school days for the various closure strategies.
Standard deviations are given in parentheses and 
$5^{th}$ and $95^{th}$ percentiles in square brackets.
Parameter values:
$\beta = 3.5 \cdot 10^{-4} \, s^{-1}$,
$\beta_{com} = 2.8 \cdot 10^{-9} \, s^{-1}$,
$1/ \mu = 2 $ day, 
$1/ \gamma = 4 $ days,
$p_A = 1/3$.
\label{table:4}
}
\end{center}}
\end{table}

\subsection*{Effect of the risk of infection in the community}
As mentioned in the Methods section,
to assess the role of the risk of infection due to contacts in the community (as opposed to those at school),
we consider a set of parameter values where $\beta_{com}$ is increased five-fold
with respect to the previous results.
That is, we use the values
$\beta = 3.5 \cdot 10^{-4} \, s^{-1}$,
$\beta_{com} = 1.4 \cdot 10^{-8} \, s^{-1}$,
$1/ \mu = 2 $ days,
$1/ \gamma = 4 $ days,
$p_A = 1/3$.
Tables \ref{table:5} and \ref{table:6} report the results we obtain
with this higher value of $\beta_{com}$ for the targeted closure strategies.
The probability of a large outbreak is much higher than in the previous case
(as shown by comparing Table~\ref{table:5} with Table~\ref{table:1}).
This probability is reduced by the targeted closure strategies, but remains comparatively large.
As observed for the smaller $\beta_{com}$, 
the decrease in the probability of a large outbreak
is larger for longer closure durations, for smaller closure-triggering thresholds,
and for closures involving more classes.
\begin{table}
{\begin{center} 
\begin{tabular}{|c|c|c|c|}
\hline
Closure strategy & Targeted class & Targeted grade & Whole school  \\
(threshold, duration)  & & & \\
\hline \hline
No closure & 65.9 & 65.9 & 65.9  \\
\hline \hline
3, 24 h & 65.0 & 62.4 & 64.1  \\
\hline
3, 48 h & 58.7 & 59.0 & 58.9 \\
\hline
3, 72 h & 58.0 & 57.4 & 53.7 \\
\hline
3, 96 h & 58.1 & 56.3 & 44.0 \\
\hline
3, 120 h & 56.4 & 51.2 & 38.9 \\
\hline
3, 144 h & 55.3 & 51.0 & 38.7  \\
\hline \hline
2, 24 h & 65.7 & 59.0 & 60.7 \\
\hline
2, 48 h & 57.3 & 56.1 & 53.2 \\
\hline
2, 72 h & 49.7 & 49.5 & 45.3 \\
\hline
2, 96 h & 46.7 & 46.2 & 43.3  \\
\hline
2, 120 h & 44.0 & 37.7 & 35.9 \\
\hline
2, 144 h & 41.5 & 36.8 & 31.7 \\
\hline
\end{tabular}
\caption{Percentage of realizations leading to an attack rate higher than $10 \%$,
for the different mitigation strategies with several closure-triggering thresholds and closure durations.
Parameter values:
$\beta = 3.5 \cdot 10^{-4} \, s^{-1}$,
$\beta_{com} = 1.4 \cdot 10^{-8} \, s^{-1}$,
$1/ \mu = 2$ days,
$1/ \gamma = 4$ days,
$p_A = 1/3$.
}
\label{table:5}
\end{center}}
\end{table}

\begin{figure}[htbp]
\begin{center}
\includegraphics[scale=0.55]{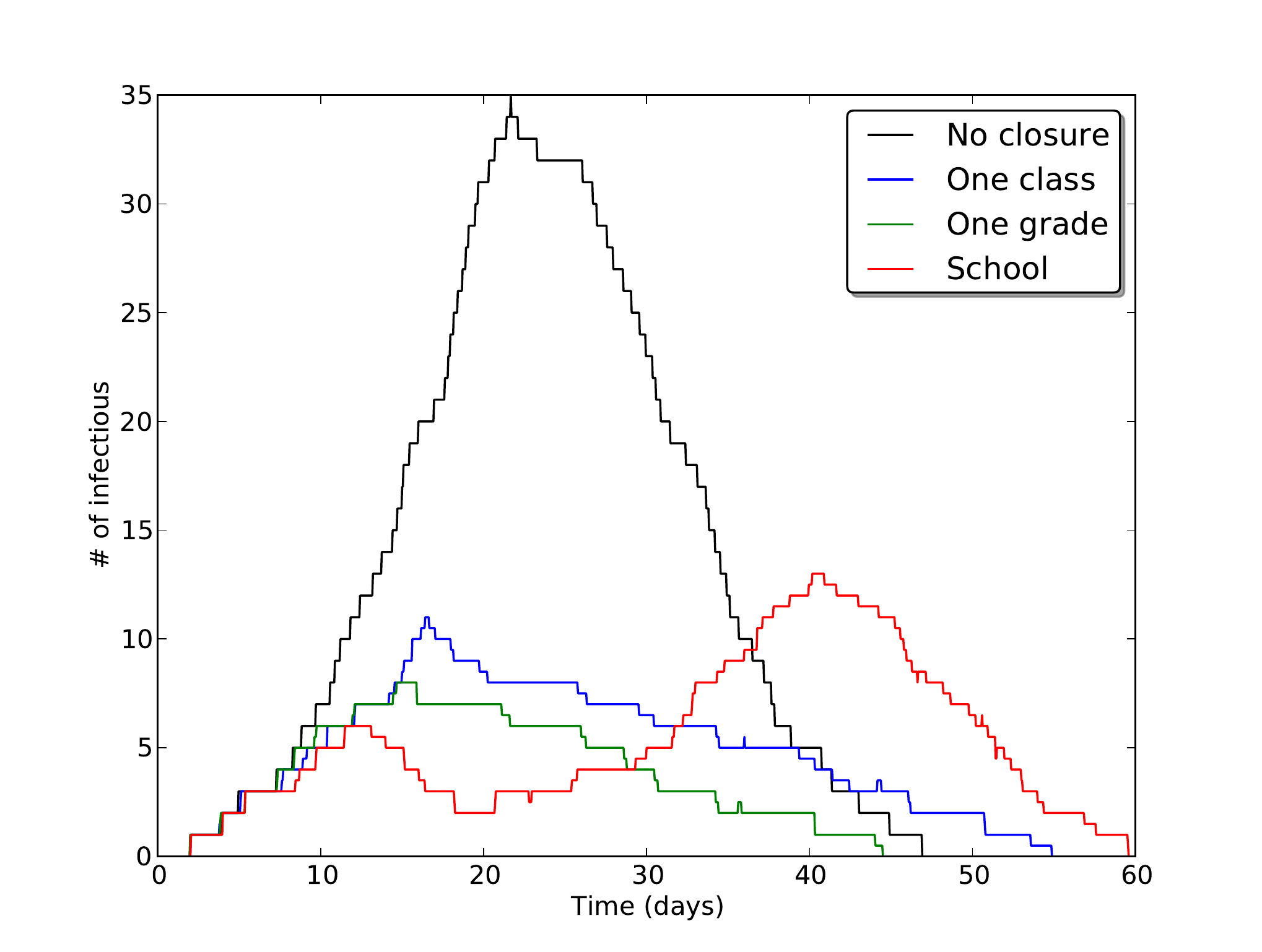}
\end{center}
\caption{Temporal evolution of the median number of infectious individuals, for
the targeted class and targeted grade closure strategies with a closure-triggering threshold
of 3 infectious individuals and a closure duration of 144 hours (that is, 6 days), compared with the scenario without closure
and the whole school closure strategy with a closure duration of 144 hours.
Here $ \beta = 3.5 \cdot 10^{-4} s^{-1} $, $ \beta_{com} = 1.4 \cdot 10^{-8} s^{-1} $, $ 1/ \mu = 2 $ days,  $ 1/ \gamma = 4 $ days, $p_A=1/3$.
Only realizations with attack rate (AR) larger than $ 10 \% $ are taken into account.
}
\label{fig:72h_3_gamma4_high}
\end{figure}

In the case of epidemics reaching more than $10\%$ of the population, however,
Table~\ref{table:6} shows that the targeted class and targeted grade closure strategies
lead to smaller attack rates than the whole school closure strategy.
Figure~\ref{fig:72h_3_gamma4_high} gives more insight into this point by showing the
epidemic curve for realizations with a final attack rate larger than 10\%, for the targeted and school closure strategies with a closure duration of 144 hours
and a closure-triggering threshold of 3 infectious individuals.  The effect of the targeted class and grade closure strategies is similar to the case of a smaller
$\beta_{com}$: these strategies lead to a smaller and shorter epidemic peak with respect to the baseline strategy. The epidemic curve for the whole
school closure strategy on the other hand is changed and has now two successive peaks; even if the first one is smaller
than for the targeted strategies, the presence of the second peak leads overall to a larger final attack rate.

\begin{table}
{\begin{center} 
\begin{tabular}{|c|c|c|c|}
\hline
Closure strategy & Targeted class & Targeted grade & Whole school  \\
(threshold, duration)  & & & \\
\hline \hline
No closure & 187 [161,192] & 187 [161,192] & 187 [161,192]  \\
\hline \hline
3, 24 h & 173 [142,196] & 176 [155,200] & 181 [153,200]  \\
\hline
3, 48 h & 155 [114,185] & 157 [69,191] & 178 [143,203] \\
\hline
3, 72 h & 131 [52,170] & 137 [40,188] & 169 [37,200] \\
\hline
3, 96 h & 118 [100,182] & 120 [28,182] & 161 [30,200]  \\
\hline
3, 120 h & 103 [35,164] & 100 [29,174] & 145 [28,196] \\
\hline
3, 144 h & 102 [39,161] & 89 [28,173] & 126 [26,196]  \\
\hline \hline
2, 24 h & 176 [145,201] & 177 [149,202] & 183 [160,203] \\
\hline
2, 48 h & 151 [89,186] & 158 [89,196] & 180 [155,201] \\
\hline
2, 72 h & 118 [37,177] & 136 [32,189] & 179 [149,201] \\
\hline
2, 96 h & 111 [30,180] & 120 [31,191] & 176 [141,202]  \\
\hline
2, 120 h & 102 [27,168] & 106 [27,188] & 176 [155,198] \\
\hline
2, 144 h & 93 [28,167] & 94 [27,185] & 174 [88,205]  \\
\hline
\end{tabular}
\caption{Average final number of cases for realizations
leading to an attack rate higher than $10\%$, for different mitigation strategies.
In square brackets we provide the $ 5^{th} $ and $ 95^{th} $ percentiles.
Parameter values:
$\beta = 3.5 \cdot 10^{-4} \, s^{-1}$,
$\beta_{com} = 1.4 \cdot 10^{-8} \, s^{-1}$,
$1/ \mu = 2 $ days,
$1/ \gamma = 4 $ days,
$p_A = 1/3$.
\label{table:6}
}
\end{center}}
\end{table}

\section{Discussion}
Since the contacts of children at school play an important role
for the propagation of many infectious diseases in the community,
it is crucial to devise efficient and cost-effective mitigation strategies
as an alternative to the closure of whole schools,
whose socio-economic costs are often considered excessive.
Inspired by the evidence that, inside one school, children do not mix homogeneously
but rather spend much more time in contact with their classmates and with other children from the same age,
we have designed targeted closure strategies at the class or grade level,
that are reactively triggered when symptomatic cases are detected.
We have simulated the dynamics of epidemic spread among school children
by using an SEIR model on top of a high-resolution time-resolved contact network measured in a primary school.
The model included asymptomatic individuals and a generic risk of infection due to random contacts
with the community when children are not at school.
Using this model we have studied the targeted strategies for class and grade closure
both in terms of their ability to mitigate the epidemic
and in terms of their impact on the schooling system, measured by the number of cancelled days of class.

%
All targeted strategies lead to an important reduction in the probability
of an outbreak reaching a large fraction of the population.
In the case of large outbreaks, targeted strategies significantly reduce
the median number of individuals affected by the epidemic.
The reduction is stronger if the strategies are triggered
by a smaller number of symptomatic cases, and if longer closures durations are used.
While the closure of one class yields a smaller mitigation effect than the closure of the whole school,
the closure of the corresponding grade (two classes) leads to a reduction of large outbreak probability
and a reduction of epidemic size that are similar to those obtained by closing the entire school,
at a much smaller cost in terms of lost class days.
In the case of large outbreaks and large risk of infection in the community,
whole-school closure might even lead to a smaller mitigation effect than targeted grade closure,
as more susceptible children would spend more time in the community,
acquiring the infection and subsequently bringing it back into the school upon re-opening.

A few important points need to be stressed.
First, the reactive character of all strategies we studied,
which are triggered by the detection of symptomatic individuals,
limits the impact on the schooling system with respect to a closure of schools
scheduled in a top-down fashion by public health authorities:
the latter would be enforced even for schools that are free of infectious individuals.
Second, targeted grade closure has in all cases a much lighter burden,
in terms of lost class days, than whole-school closure.  
Given also its good performance in the mitigation of outbreaks,
it thus represents an interesting alternative strategy. 
Finally, we recall that grade closure corresponds
to closing the class in which symptomatic children are detected
and the class which has the most contacts with it.
To assess this relation between classes,
we do not need the very detailed knowledge of the contact patterns we used in this study:
rather, readily available information such as class schedules and classroom locations~\cite{Smieszek:2013}
may be sufficient to retrieve this information.
This has important public health consequences,
as it implies that the targeted mitigation strategies studied here
might actually be carried out in the general case, without high-resolution contact network data.

%
Some limitations of this study are worth mentioning.
While the strategies we discussed could be designed and implemented with limited information,
they were only tested using one specific dataset corresponding to one particular school.
The high-resolution contact network data we used only spans two days of school activity,
and had to be extended  longitudinally by using a repetition procedure.
This technique is commonly used to simulate epidemic spread on temporal data,
but it does introduce strong temporal correlations in the extended dataset
and it may fail to correctly model the day-to-day heterogeneity of contact patterns~\cite{Stehle:2011b}.
While variations in the repetition of contacts from one day to the next are known to modify the attack rate
of an epidemic~\cite{Smieszek:2009}, we expect that the relative efficiency of the strategies we considered
should be robust with respect to other temporal extensions strategies~\cite{Stehle:2011b}.
Moreover, this limitation should be less of an issue in school settings,
where mixing patterns are shaped by a regular activity schedule and have a strong periodic character.
Another limitation of this study is the simplistic coupling with the community that we used:
our high-resolution contact network does not include contacts happening outside of school,
so we introduced in our model a free parameter that describes a generic risk of transmission from the community.
Even though our results are robust with respect to important variations in this parameter,
it would be desirable to inform the model with empirical data on the contacts that children have
with members of the community, or with one another outside of school.

%
The limitations described above point to several directions for further research.
It would be interesting to validate the targeted class and grade closure strategies
using high-resolution data describing the contacts of children in other schools and over longer timescales,
if such dataset become available in the future.
In particular it would be interesting to consider larger schools,
for which the closure of more than two classes may represent a valid intervention with intermediate impact.
High-resolution measurement of contact patterns within a school could be coupled
with surveys administered to the same children, to estimate their contact rates off-school
and to model their contacts with other individuals of different age classes in the community.
Such data could be used to refine the model used here,
but also to design agent-based models at a larger scale (e.g., urban or geographic),
spanning several schools. This would allow to generalize the strategies introduced
in this paper to the case of multiple schools, and to evaluate their relative efficiency,
in particular comparing targeted strategies with the general closure of all schools
in the relevant geographical region. 
Finally, we have discussed how targeted interventions can be guided by readily-available
information on the school activities and organisational structure.
New techniques to tease apart meso-scale activity patterns in high-resolution contact data~\cite{Gauvin:2014}
could be used to design and guide targeted intervention aimed not just at closing classes
but, for example, at suspending or modifying specific activities in the school
that involve the shared use of spaces (e.g., sports activities, time in the playground,
lunch at the cafeteria, etc.)

\acknowledgments{
This work is partially supported by the Lagrange Project of the ISI
Foundation funded by the CRT Foundation to AB, CC and VG, by the Q-ARACNE project
funded by the Fondazione Compagnia di San Paolo to CC, 
by the HarMS-flu project (ANR-12-MONU-0018) funded by the French ANR to AB,
and by the FET Multiplex Project (EU-FET-317532) funded by the European Commission
to AB and CC.}


\newpage

\section*{Mitigation of infectious disease at school: \\
targeted class closures vs school closures \\
Supplementary Text}

\subsection*{Results of the spreading simulations for modified SEIR model parameters}

As mentioned in the main text, we report here the results of numerical simulations performed with different sets of parameters for the SEIR epidemic model,
in order to assess the robustness of the obtained results. We use here 
\begin{itemize}
\item[(i)] $ \beta = 6.9 \cdot 10^{-4} s^{-1} $, $ \beta_{com} = 2.8 \cdot 10^{-9} s^{-1} $, $ 1/ \mu = 1 $ day,  $ 1/ \gamma = 2 $ days
\item[(ii)] $ \beta = 1.4 \cdot 10^{-3} s^{-1} $,  $ \beta_{com} = 2.8 \cdot 10^{-9} s^{-1} $,  $ 1/ \mu = 0.5 $ day; $ 1/ \gamma = 1 $ day, 
 \end{itemize}
 and the fraction of infected asymptomatic individuals is $p_A = 1/3 $ in both cases.
 As the average latent and infectious periods are shorter than for the simulations presented in the main text, the epidemic will unfold here on shorter timescales.
 
 We implement the same mitigation strategies as in the main text, and, in order to compare their relative efficiencies, we 
 focus on the fraction of stochastic realizations that yield a global attack rate higher than $ 10 \% $, on 
the final average number of cases and on the temporal evolution of the number of infectious individuals. We 
 compare the results obtained by implementing the various mitigation measures with the baseline case, represented by the situation in which the only mitigation 
 strategy consists in the isolation of the symptomatic individuals once they are detected.

\subsubsection*{Results}

In Table \ref{table:S1} and  \ref{table:S1bis}  we show the fraction of stochastic realizations leading to an attack rate (AR) higher than $ 10 \% $, for various mitigation 
measures, compared with the same result obtained when no closure was implemented. 
As found for the set of parameters used in the main text's results, the reduction of the probability of a severe outbreak is important 
for all strategies, and increases for smaller triggering thresholds and longer closure durations. The closure of the whole school
always leads to the most important effect, but the grade closure has as well a strong impact on the probability of occurrence of large outbreaks.
 
\begin{table}
{\begin{center} 
\begin{tabular}{|c|c|c|c|}
\hline
Closure strategy & Targeted class & Targeted grade & Whole school  \\
(Threshold, duration)  & & & \\
\hline \hline
No closure & 26.4 & 26.4 & 26.4  \\
\hline \hline
3, 24 h & 22.8 & 21.8 & 21.7  \\
\hline
3, 48 h & 21.8 & 17.0 & 13.3 \\
\hline
3, 72 h & 18.6 & 15.8 & 6.3 \\
\hline
3, 96 h & 16.6 & 14.1 & 5.4  \\
\hline \hline
2, 24 h & 22.2 & 21.5 & 19.7 \\
\hline
2, 48 h & 16.8 & 17.0 & 12.6 \\
\hline
2, 72 h & 11.9 & 11.2 & 3.4 \\
\hline
2, 96 h & 10.2 & 8.7 & 3.5  \\
\hline
\end{tabular}
\caption{Percentage of realizations leading to an attack rate higher than $ 10 \% $, for the various mitigation strategies with various thresholds and closure durations.
$ \beta = 6.9 \cdot 10^{-4} s^{-1} $, $ \beta_{com} = 2.8 \cdot 10^{-9} s^{-1} $, $ 1/ \mu = 1 $ day,  $ 1/ \gamma = 2 $ days, $p_A=1/3$.
}
\label{table:S1}
\end{center}}
\end{table}
\begin{table}
{\begin{center} 
\begin{tabular}{|c|c|c|c|}
\hline
Closure strategy & Targeted class & Targeted grade & Whole school  \\
(Threshold, duration)  & & & \\
\hline \hline
No closure & 13.4 & 13.4 & 13.4  \\
\hline \hline
3, 24 h & 10.0 & 8.6 & 6.9  \\
\hline
3, 48 h & 7.3 & 6.7 & 3.2 \\
\hline 
3, 72 h & 8.8 & 5.9 & 4.3 \\
\hline \hline
2, 24 h & 7.4 & 6.0 & 4.4 \\
\hline
2, 48 h & 5.6 & 3.1 & 2.3 \\
\hline
2, 72 h & 5.3 & 3.2 & 1.9 \\
\hline
\end{tabular}
\caption{Percentage of realizations leading to an attack rate higher than $ 10 \% $, for the various mitigation strategies with various thresholds and closure durations.
Here $ \beta = 1.4 \cdot 10^{-3} s^{-1} $, $ \beta_{com} = 2.8 \cdot 10^{-9} s^{-1} $, $ 1/ \mu = 0.5 $ day,  $ 1/ \gamma = 1 $ day, $p_A=1/3$.
}\label{table:S1bis}
\end{center}}
\end{table}

Tables \ref{table:S2} and  \ref{table:S2bis} give 
the final number of cases for both parameter sets, computed for the realisations leading to an attack rate larger than $10\%$ for the various mitigation strategies.
All strategies lead to a reduction in the final number of cases. Interestingly, and similarly to the case shown in the main text,
this reduction is very similar for the various closures (class, grade or whole school) if the closure triggering threshold
is small enough: in cases of large outbreaks, closing only one class is as effective as closing the whole school. As in the main text, we also note that large confidence 
intervals are observed, limiting the predictability of the final number of cases. Finally, as the duration of the closures is increased, the impact
on the spread saturates at a value close to the sum of the latent and infectious periods, as in the main text.

\begin{table}
{\begin{center} 
\begin{tabular}{|c|c|c|c|}
\hline
Closure strategy & Targeted class & Targeted grade & Whole school  \\
(Threshold, duration)  & & & \\
\hline \hline
No closure & 154 [106,209] & 154 [106,209] & 154 [106,209]  \\
\hline \hline
3, 24 h & 133 [91,194] & 131 [90,192] & 141 [101,206]  \\
\hline
3, 48 h & 100 [68,175] & 99 [68,174] & 111 [72,179] \\
\hline
3, 72 h & 71 [39,135] & 63 [37,133] & 57 [35,129] \\
\hline
3, 96 h & 66 [38,124] & 58 [37,122] & 38 [21,89]  \\
\hline \hline
2, 24 h & 130 [88,192] & 126 [83,189] & 141 [103,201] \\
\hline
2, 48 h & 94 [67,169] & 91 [66,169] & 124 [99,197] \\
\hline
2, 72 h & 54 [36,120] & 52 [36,119] & 53 [34,126] \\
\hline
2, 96 h & 51 [37,109] & 45 [35,104] & 43 [33,107]  \\
\hline
\end{tabular}
\caption{Average final number of cases, computed for the realisations leading to an attack rate larger than $10 \% $ for the various mitigation strategies; 
the brackets provide the $ 5^{th} $ and $ 95^{th} $ percentiles.
$ \beta = 6.9 \cdot 10^{-4} s^{-1} $, $ \beta_{com} = 2.8 \cdot 10^{-9} s^{-1} $, $ 1/ \mu = 1 $ day,  $ 1/ \gamma = 2 $ days, $p_A=1/3$.
}
\label{table:S2}
\end{center}}
\end{table}
\begin{table}
{\begin{center} 
\begin{tabular}{|c|c|c|c|}
\hline
Closure strategy & Targeted class & Targeted grade & Whole school  \\
(Threshold, duration)  & & & \\
\hline \hline
No closure & 61 [26,128] & 61 [26,128] & 61 [26,128]  \\
\hline \hline
3, 24 h & 53 [26,98] & 51 [27,100] & 55 [26,101]  \\
\hline
3, 48 h & 45 [26,88] & 43 [26,68] & 39 [25,70] \\
\hline
3, 72 h & 45 [26,86] & 41 [26,67] & 34 [25,48] \\
\hline \hline
2, 24 h & 48 [25,85] & 47 [26,81] & 51 [26,104] \\
\hline
2, 48 h & 39 [25,64] & 38 [26,63] & 41 [26,81] \\
\hline
2, 72 h & 38 [26,64] & 36 [25,51] & 35 [25,48] \\
\hline
\end{tabular}
\caption{Average final number of cases, computed for the realisations leading to an attack rate larger than $10 \%$, for the various mitigation strategies; 
the brackets provide the $ 5^{th} $ and $ 95^{th} $ percentiles.
Here $ \beta = 1.4 \cdot 10^{-3} s^{-1} $, $ \beta_{com} = 2.8 \cdot 10^{-9} s^{-1} $, $ 1/ \mu = 0.5 $ day,  $ 1/ \gamma = 1 $ day, $p_A=1/3$.
}\label{table:S2bis}
\end{center}}
\end{table}

\newpage 

\begin{figure}[htbp]
\begin{center}
\includegraphics[scale=0.55]{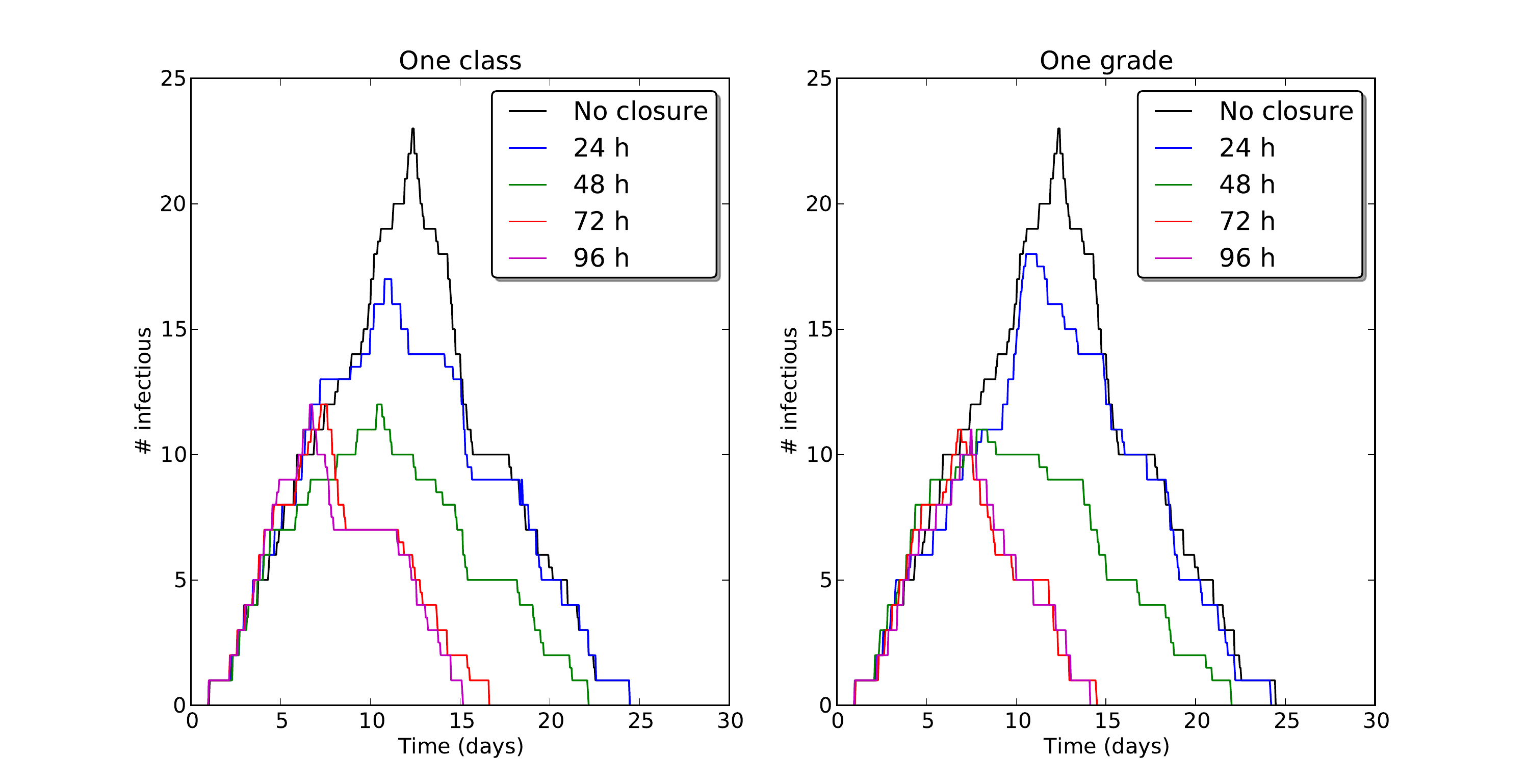}
\includegraphics[scale=0.45]{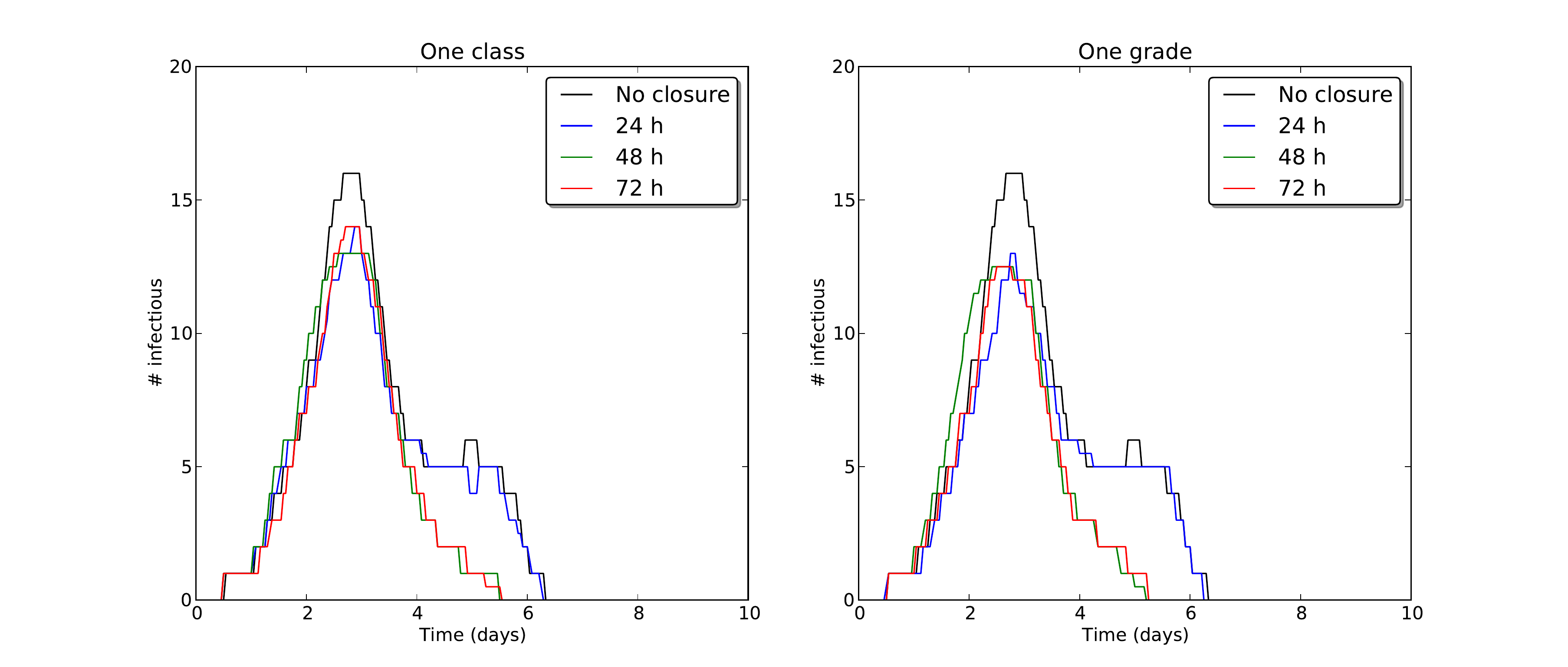}
\end{center}
\caption{Temporal evolution of the number of infectious individuals, for various closure durations, 
in the case of a closure-triggering threshold equal to 3. Left: targeted class closure; right: targeted grade closure. 
Top plots: $ \beta = 6.9 \cdot 10^{-4} s^{-1} $, $ \beta_{com} = 2.8 \cdot 10^{-9} s^{-1} $, $ 1/ \mu = 1 $ day,  $ 1/ \gamma = 2 $ days, $p_A=1/3$.
Bottom plots: $ \beta = 1.4 \cdot 10^{-3} s^{-1} $,  $ \beta_{com} = 2.8 \cdot 10^{-9} s^{-1} $, $ 1/ \mu = 0.5 $ day,  $ 1/ \gamma = 1 $ day, $p_A=1/3$.
Only runs with attack rate (AR) larger than $ 10 \% $ are taken into account.}
\label{fig:thr3}
\end{figure}

\begin{figure}[htbp]
\begin{center}
\includegraphics[scale=0.36]{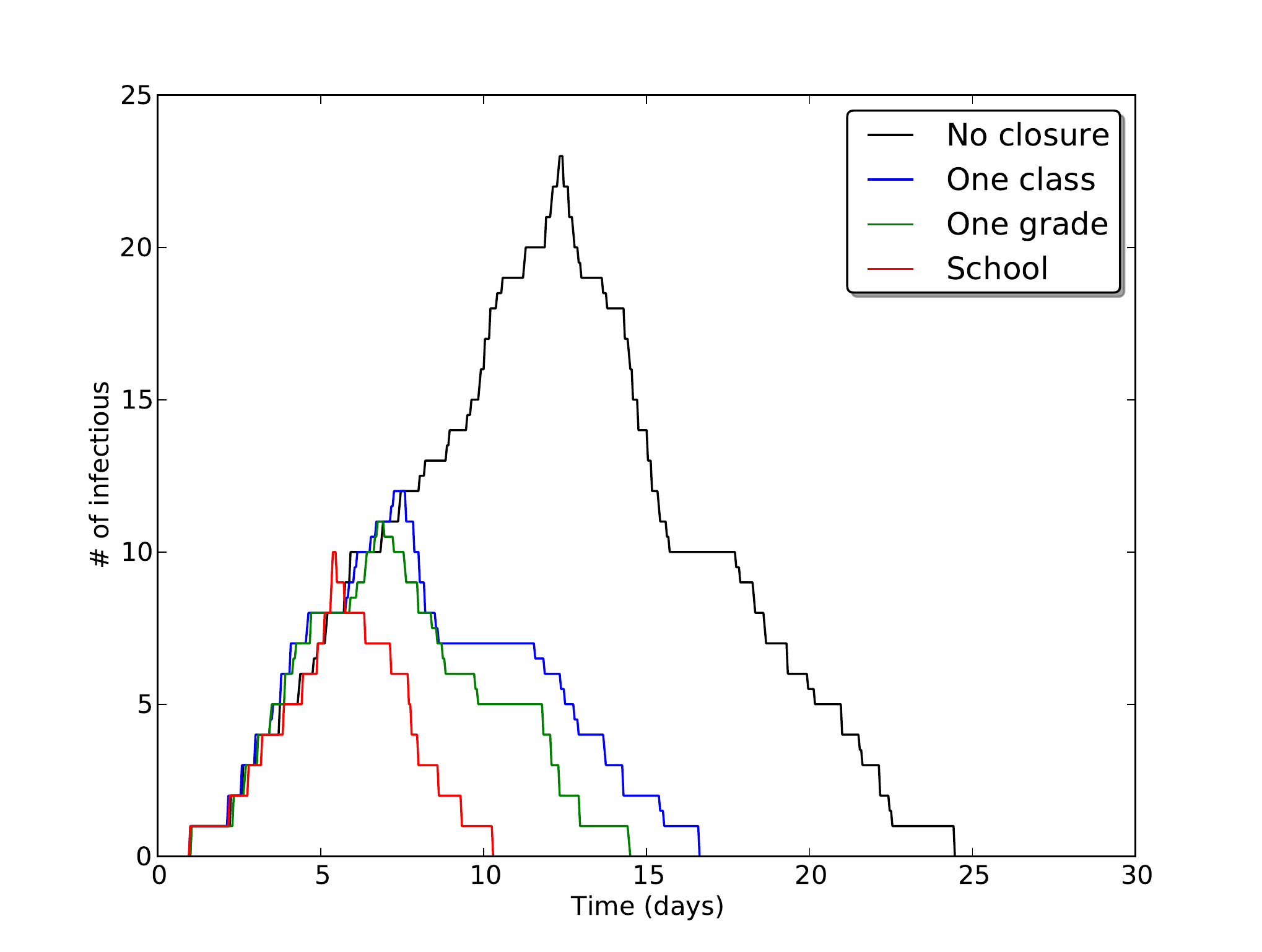}
\includegraphics[scale=0.36]{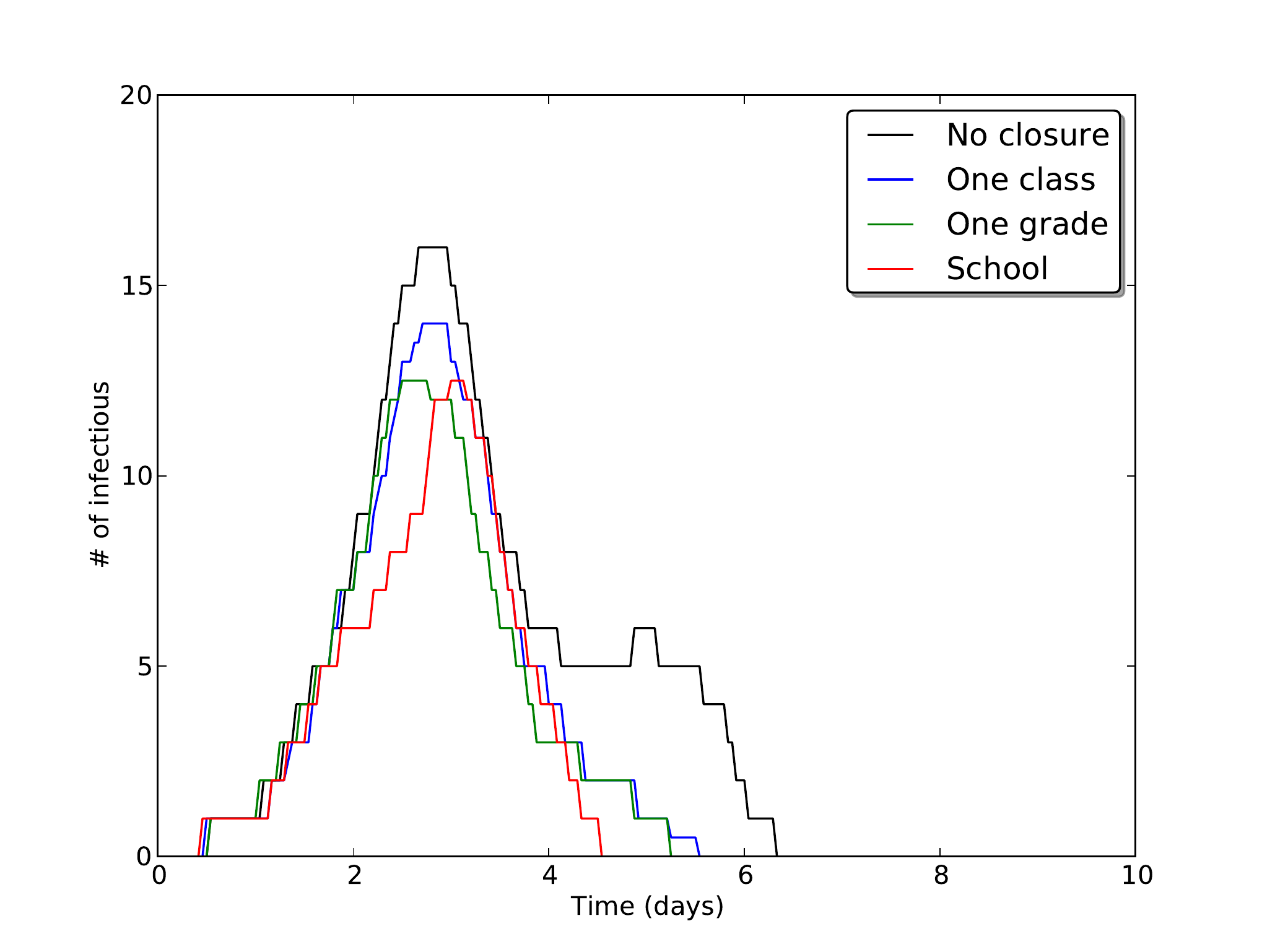}
\end{center}
\caption{Temporal evolution of the number of infectious individuals, for 
the targeted class and targeted grade closure strategies with a closure-triggering threshold
of 3 infectious individuals and a closure duration of 72 hours, compared with the scenario without closure
and the whole school closure strategy with a closure duration of 72 hours.
Left: $ \beta = 6.9 \cdot 10^{-4} s^{-1} $, $ \beta_{com} = 2.8 \cdot 10^{-9} s^{-1} $, $ 1/ \mu = 1 $ day,  $ 1/ \gamma = 2 $ days, $p_A=1/3$.
Right: $ \beta = 1.4 \cdot 10^{-3} s^{-1} $, $ \beta_{com} = 2.8 \cdot 10^{-9} s^{-1} $, $ 1/ \mu = 0.5 $ day,  $ 1/ \gamma = 1 $ day, $p_A=1/3$.
Only realizations with attack rate (AR) larger than $ 10 \% $ are taken into account.}
\label{fig:th3_72h}
\end{figure}

Figures \ref{fig:thr3} and \ref{fig:th3_72h} complement the results by showing the temporal evolution of the median 
number of infectious individuals, for realisations leading to an attack rate
larger than $10\%$. 

Figure \ref{fig:thr3}  shows the effect of various closure durations, for a closure triggering threshold of three symptomatic individuals and for the
targeted class and grade closure strategies. The epidemic curve unfolds on a shorter time-scale than in the case shown in the main text, as expected. 
As a result, the various strategies do not change much the epidemic peak timing, and have a smaller influence on the global duration of the spread,
especially for the fastest spread. 
Moreover, in both cases, very similar epidemic curves are obtained when the closure duration becomes larger than the sum of 
$1/\mu$ and $1/\gamma$, while shorter durations yield a smaller effect.  The optimal closure duration is thus close to the
sum of the latent and infectious periods.

Figure~\ref{fig:th3_72h}  finally compares
the epidemic curves for the targeted class closure, targeted grade closure, and whole school closure strategies, for
a closure duration of three days and a closure-triggering threshold of three symptomatic individuals.  
The peak heights are very similar for the targeted grade closure strategy and for the closure of the whole school.

Overall, these results represent a very similar phenomenology with respect to the parameters used in the simulations shown in the main text, indicating the
robustness of our results with respect to changes in the disease parameters and the relevance of targeted class and grade strategies as alternative measures
to the closure of whole schools. 

\newpage 
\subsection*{Comparison of targeted and random class and grade closures} 

We consider here strategies based on random closure of classes:
whenever the number of symptomatic infectious individuals detected in any class
reaches a certain threshold, 
\begin{itemize}
\item[(iv)] one random class, different from the one in which symptomatic individuals were detected, is closed (``random class closure'' strategy)
\item[(v)] this class and a randomly chosen one in a different grade are closed (``mixed class closure'' strategy).
\end{itemize}
We use here 
$ \beta = 6.9 \cdot 10^{-4} s^{-1} $, $ \beta_{com} = 2.8 \cdot 10^{-9} s^{-1} $, $ 1/ \mu = 1 $ day,  $ 1/ \gamma = 2 $ days, $p_A=1/3$.

Tables \ref{table:S3} and \ref{table:S4} compare the effect of the targeted class and grade closure with the partially random closure
of one or two classes. Closing one class chosen at random (different than the one in which the infectious
individuals are detected) leads only to a marginal decrease in the probability to 
obtain an attack rate higher than $ 10 \% $ and in the number of individuals affected by large spreads.

\begin{table}
{\begin{center} 
\begin{tabular}{|c|c|c|c|c|}
\hline
Closure strategy &  Targeted class & Random class &  Targeted grade & Mixed  \\
(Threshold, duration)  & & & & \\
\hline \hline
No closure & 26.4 & 26.4 & 26.4 & 26.4 \\
\hline \hline
3, 24 h & 22.8 & 26.2 & 21.8 & 22.1 \\
\hline
3, 48 h & 21.8 & 24.4 & 17.0 & 18.1 \\
\hline
3, 72 h & 18.6 & 26.0 & 15.8 & 17.6 \\
\hline
3, 96 h & 16.6 & 23.2 & 14.1 & 16.5 \\
\hline \hline
2, 24 h & 22.2 & 23.8 & 21.5 & 22.2 \\
\hline
2, 48 h & 16.8 & 24.3 & 17.0 & 17.9 \\
\hline
2, 72 h & 11.9 & 24.8 & 11.2 & 14.4 \\
\hline
2, 96 h & 10.2 & 23.8 & 8.7 & 13.1 \\
\hline
\end{tabular}
\caption{Percentage of runs leading to an attack rate larger than $ 10 \% $ for the targeted
and random closure strategies.
$ \beta = 6.9 \cdot 10^{-4} s^{-1} $, $ \beta_{com} = 2.8 \cdot 10^{-9} s^{-1} $, $ 1/ \mu = 1 $ day,  $ 1/ \gamma = 2 $ days, $p_A=1/3$.
}
\label{table:S3}
\end{center}}
\end{table}

In the mixed strategy on the other hand, the class in which the infectious individuals have been detected is closed, and a second one, 
chosen at random in a different grade, is closed as well. This leads to an effect almost as strong as the targeted grade closure strategy.

\begin{table}
{\begin{center} 
\begin{tabular}{|c|c|c|c|c|}
\hline
Closure strategy & Targeted class & Random class & Targeted grade& Mixed \\
(Threshold, duration)  & & & & \\
\hline  \hline
No closure & 154 [106,209] & 154 [106,209] & 154 [106,209] & 154 [106,209] \\
\hline \hline
3, 24 h & 133 [90,194] & 151 [106,207] & 131 [90,192] & 132 [90,195] \\
\hline
3, 48 h & 100 [68,175] & 145 [103,201] & 99 [68,174] & 102 [70,176] \\
\hline
3, 72 h & 71 [39,135] & 130 [89,193] & 63 [37,133] & 64 [37,135] \\
\hline
3, 96 h & 66 [38,124] & 123 [87,191] & 58 [37,122] & 63 [37,133] \\
\hline \hline
2, 24 h & 130 [88,192] & 153 [105,207] & 126 [83,189] & 128 [84,192] \\
\hline
2, 48 h & 94 [67,169] & 150 [104,207] & 91 [66,169] & 95 [67,170] \\
\hline
2, 72 h & 54 [36,120] & 135 [93,190] & 52 [36,119] & 56 [37,123] \\
\hline
2, 96 h & 51 [37,109] & 119 [84,189] & 45 [35,104] & 48 [35,106] \\
\hline
\end{tabular}
\label{tab:randomAR}
\caption{Average final number of cases of the spread, for realisations with $ AR > 10 \% $, for the targeted
and random strategies. 
$ \beta = 6.9 \cdot 10^{-4} s^{-1} $, $ \beta_{com} = 2.8 \cdot 10^{-9} s^{-1} $, $ 1/ \mu = 1 $ day,  $ 1/ \gamma = 2 $ days, $p_A=1/3$.
The brackets give the $ 5^{th} $ and $ 95^{th} $ percentiles.}
\label{table:S4}
\end{center}}
\end{table}

\begin{figure}[htbp]
\begin{center}
\includegraphics[scale=0.43]{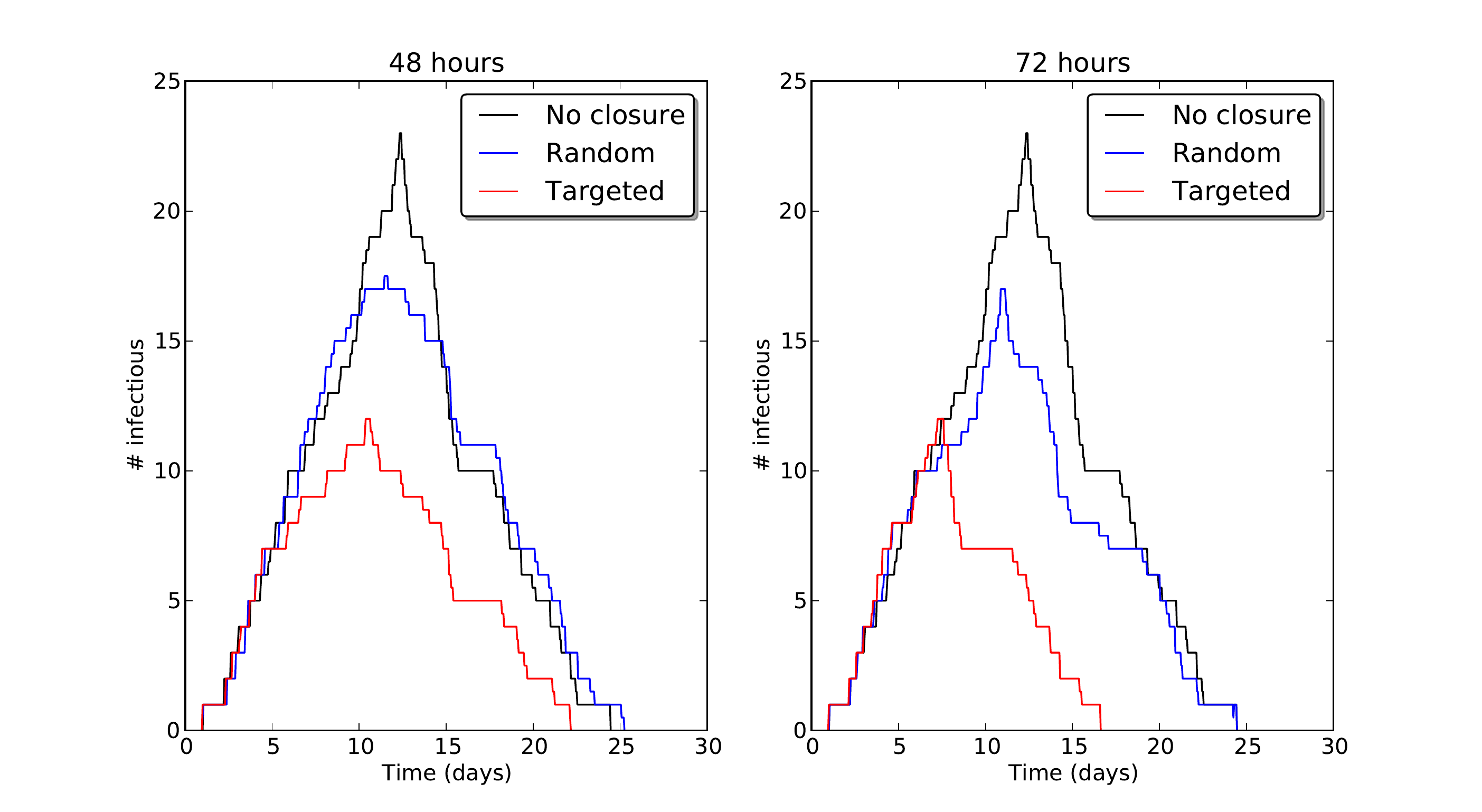}
\end{center}
\caption{Temporal evolution of the number of infectious individuals for targeted and random class closure strategies, 
for a closure-triggering threshold of 3 infectious individuals, compared with the scenario in which no closure is implemented.
The duration of the class closures is 48 hours (Left) and 72 hours (right). 
$ \beta = 6.9 \cdot 10^{-4} s^{-1} $, $ \beta_{com} = 2.8 \cdot 10^{-9} s^{-1} $, $ 1/ \mu = 1 $ day,  $ 1/ \gamma = 2 $ days, $p_A=1/3$.
Only
realizations with attack rate (AR) larger than $ 10 \% $ are taken into account.}
\label{fig:random_3}
\end{figure}

\begin{figure}[htbp]
\begin{center}
\includegraphics[scale=0.43]{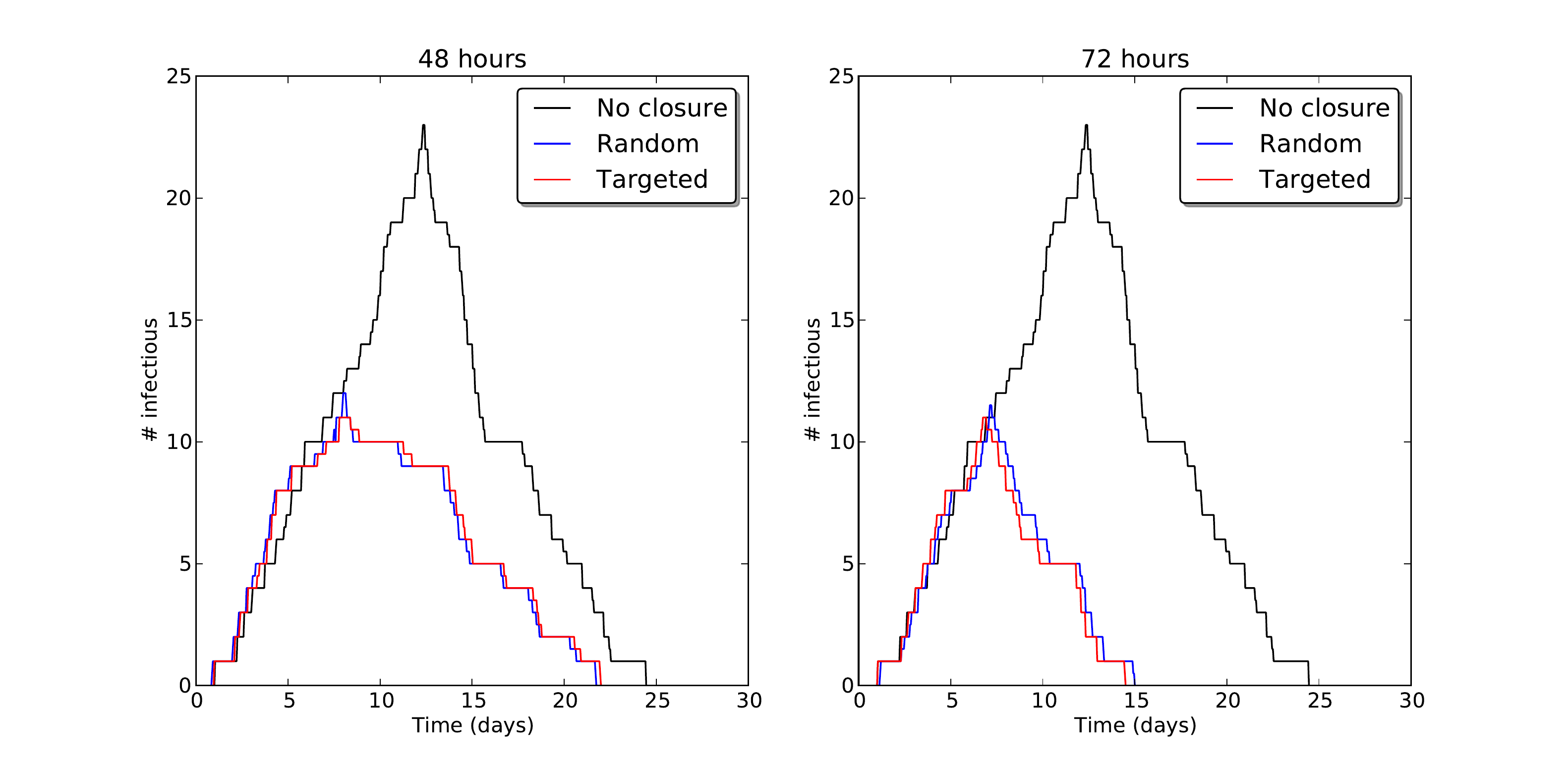}
\end{center}
\caption{Temporal evolution of the number of infectious individuals for the targeted grade closure and the mixed closure strategies,
for a closure-triggering threshold of 3 infectious individuals, compared with the scenario in which no closure is implemented.
The duration of the class closures is 48 hours (Left) and 72 hours (right). In the mixed
closure strategy, the class in which the infectious individuals are detected is closed, as well as a second class from a different 
grade. $ \beta = 6.9 \cdot 10^{-4} s^{-1} $, $ \beta_{com} = 2.8 \cdot 10^{-9} s^{-1} $, $ 1/ \mu = 1 $ day,  $ 1/ \gamma = 2 $ days, $p_A=1/3$.
Only realizations with attack rate (AR) larger than $ 10 \% $ are taken into account.}
\label{fig:random_seconda_3}
\end{figure}

Figures  \ref{fig:random_3} and \ref{fig:random_seconda_3} report the temporal evolution of the median number of infectious 
individuals in the targeted and random strategies, leading to the same conclusion: the targeted closure of a class leads to a much smaller peak than the closure of a random class; the mixed strategy leads on the other hand to an epidemic curve that is very close to the case of a targeted grade closure.

\newpage

\end{document}